\documentclass[pre,nofootinbib,twocolumn,showpacs,showkeys,preprintnumbers,floatfix]{revtex4-1}

\usepackage{hyperref}
\usepackage{etex}
\usepackage{ifpdf}
\usepackage{hyperref}
\usepackage{dcolumn}
\usepackage{url}
\usepackage{amsmath}
\usepackage{amscd}
\usepackage{amsfonts}
\usepackage{amssymb}
\usepackage{amsthm}
\usepackage{xfrac}
\usepackage{graphicx}
\usepackage{braket}
\usepackage{bm}   % bold math
\usepackage{bbm}
\usepackage{verbatim}
\usepackage{stmaryrd}
\usepackage{xcolor}
\usepackage{setspace}
\usepackage{diagbox}
\usepackage{tabulary}
\usepackage{mathtools}
\usepackage{mathrsfs}
\usepackage{placeins}
\usepackage{xfrac}
\usepackage{tikz}
\usetikzlibrary{arrows}
\usetikzlibrary{automata}

\hypersetup{colorlinks=true, linkcolor=blue, citecolor=blue, pagebackref=true}
\RequirePackage[hyperpageref]{backref}

\newcommand{\MeasAlphabet}  {\mathcal{A}}
\newcommand{\CausalState}   { \mathcal{S} }

\newcommand{\ProcessAlphabet}   {\MeasAlphabet}

\newcommand{\forward}{+}
\newcommand{\reverse}{-}
\newcommand{\forwardreverse}{\pm} % \pm

\newcommand{\FutureCausalState} { {\CausalState}^{\forward} }

\newcommand{\PastCausalState}   { {\CausalState}^{\reverse} }

\newcommand{\lastindex}[2]{
  \edef\tempa{0}
  \edef\tempb{#2}
  \ifx\tempa\tempb
    \edef\tempc{#1}
  \else
    \edef\tempa{0}
    \edef\tempb{#1}
    \ifx\tempa\tempb
      \edef\tempc{#2}
    \else
      \edef\tempc{#1+#2}
    \fi
  \fi
  \tempc
}

\newcommand{\I}{\mathbf{I}}

\newcommand{\CSjoint}[1][,]{
   \edef\tempa{:}
   \edef\tempb{#1}
   \ifx\tempa\tempb
      \ensuremath{\FutureCausalState\!#1\PastCausalState}
   \else
      \ensuremath{\FutureCausalState#1\PastCausalState}
   \fi
}

\newif\ifpm
\edef\tempa{\forwardreverse}
\edef\tempb{\pm}
\ifx\tempa\tempb
   \pmfalse
\else
   \pmtrue
\fi

\renewcommand{\H}{\operatorname{H}}
\renewcommand{\I}{\operatorname{I}}
\newcommand{\kB} { k_\text{B} }

\newcommand{\Lhat} { \widehat{L} }

\theoremstyle{plain}    
\theoremstyle{plain}    
\theoremstyle{plain}    
\theoremstyle{plain}    
\theoremstyle{plain}    
\theoremstyle{plain}    
\theoremstyle{plain}    
\theoremstyle{plain}    
\theoremstyle{plain}    
\theoremstyle{plain}    
\theoremstyle{plain}    
\theoremstyle{plain}    
\theoremstyle{plain}    

\def\clap#1{\hbox to 0pt{\hss#1\hss}}

\begin{document}

\title{
%Heat Dissipation and Energy Consumption\\
%When Generating Random Numbers\\
%Or\\
Thermodynamics of Random Number Generation}

\author{Cina Aghamohammadi}
\email{caghamohammadi@ucdavis.edu}

\author{James P. Crutchfield}
\email{chaos@ucdavis.edu}

\affiliation{Complexity Sciences Center and Department of Physics,\\
University of California at Davis, One Shields Avenue, Davis, CA 95616}

\date{\today}
\bibliographystyle{unsrt}

% ************************* ABSTRACT *************************
\begin{abstract}
We analyze the thermodynamic costs of the three main approaches to
generating random numbers via the recently introduced Information Processing Second Law. Given access to a specified source of randomness, a random number
generator (RNG) produces samples from a desired target probability
distribution. This differs from pseudorandom number generators (PRNG) that use
wholly deterministic algorithms and from true random number generators (TRNG)
in which the randomness source is a physical system. For each class, we analyze
the thermodynamics of generators based on algorithms implemented as
finite-state machines, as these allow for direct bounds on the required
physical resources. This establishes bounds on heat dissipation and work
consumption during the operation of three main classes of RNG
algorithms---including those of von Neumann, Knuth and Yao, and Roche and
Hoshi---and for PRNG methods. We introduce a general TRNG and determine its
thermodynamic costs exactly for arbitrary target distributions. The results
highlight the significant differences between the three main approaches to
random number generation: One is work producing, one is work consuming, and the
other is potentially dissipation neutral. Notably, TRNGs can both generate
random numbers and convert thermal energy to stored work. These thermodynamic
costs on information creation complement Landauer's limit on the irreducible
costs of information destruction.
\end{abstract}


\preprint{Santa Fe Institute Working Paper 16-12-XXX}
\preprint{arxiv.org:1612.XXXXX [cond-mat.stat-mech]}

\maketitle 
% ****************************************************************

%\tableofcontents
\setstretch{1.1}

\section{Introduction} 

Random number generation is an essential tool these days in simulation and
analysis. Applications range from statistical sampling \cite{Coch07a},
numerical simulation \cite{Jerr84a}, cryptography \cite{Stin05a}, program
validation \cite{Sarg05aa}, and numerical analysis \cite{Stoe13a} to machine
learning \cite{Alpa14a} and decision making in games \cite{Conw76a} and in
politics \cite{Dowl15a}. More practically, a significant fraction of all the
simulations done in physics \cite{Rubi11a} employ random numbers to greater or
lesser extent.

Random number generation has a long history, full of deep design challenges and
littered with pitfalls. Initially, printed tables of random digits were used
for scientific work, first documented in 1927 \cite{Knut81a}. A number of
analog physical systems, such as reversed-biased Zener diodes \cite{Motc73a}
or even Lava\textsuperscript{\textregistered} Lamps \cite{Mend96a}, were also
employed as sources of randomness; the class of so-called \emph{noise generators}. One of the first digital machines that
generated random numbers was built in 1939 \cite{Kend38a}. With the advent of
digital computers, analog methods fell out of favor, displaced by a growing
concentration on arithmetical methods that, running on deterministic digital
computers, offered flexibility and reproducibility. An early popular approach
to digital generation was the \emph{linear congruential method} introduced in
1950 \cite{Lehm51a}. Since then many new arithmetical methods have been
introduced \cite{Wich82a,Blum86a,Masc95a,Kels99a,Mars03a,Salm11a}.

The recurrent problem in all of these strategies is demonstrating that the
numbers generated were, in fact, random. This concern eventually lead to
Chaitin's and Kolmogorov's attempts to find an algorithmic foundation for
probability theory \cite{Kolm65,Chai66,Mart66a,Levi74a,Vita93a,Kolm83}. Their
answer was that an object is random if it cannot be compressed: random objects
are their own minimal description. The theory exacts a heavy price, though:
identifying randomness is uncomputable \cite{Vita93a}.

Despite the formal challenges, many physical systems appear to behave
randomly.  Unstable nuclear decay processes obey Poisson statistics
\cite{Knol10a}, thermal noise obeys Gaussian statistics \cite{Dave58a}, cosmic
background radiation exhibits a probabilistically fluctuating temperature field
\cite{Yosh01a}, quantum state measurement leads to stochastic outcomes
\cite{Jenn00a,Stef00a,Acin16a}, and fluid turbulence is governed by an
underlying chaotic dynamic \cite{Bran83}. When such physical systems are used
to generate random numbers one speaks of \emph{true random number generation}
\cite{Stip14a}.

Generating random numbers without access to a source of randomness---that is,
using arithmetical methods on a deterministic finite-state machine, whose logic
is physically isolated---is referred to as \emph{pseudorandom number
generation}, since the numbers must eventually repeat and so, in principle, are
not only not random, but are exactly predictable \cite{Gent13,Rubi98a}. John
von Neumann was rather decided about the pseudo-random distinction: ``Any one
who considers arithmetical methods of producing random digits is, of course, in
a state of sin'' \cite{Vneu51}. Nonetheless, these and related methods dominate
today and perform well in many applications.

Sidestepping this concern by assuming a given source of randomness,
\emph{random number generation} (RNG) \cite{Devr86a} is a complementary problem
about the transformation of randomness: Given a specific randomness source,
whose statistics are inadequate somehow, how can we convert it to a source that
meets our needs? And, relatedly, how efficiently can this be done?

Our interest is not algorithmic efficiency, but thermodynamic efficiency, since
any practical generation of random numbers must be physically embedded. What
are the energetic costs---energy dissipation and power inputs---to harvest a
given amount of information? This is a question, at root, about a particular
kind of information processing---viz., information creation---and the demands
it makes on its physical substrate. In this light, it should be seen as exactly
complementary to Landauer's well known limit on the thermodynamic costs of
information destruction (or erasure) \cite{Land61a,Benn82}.

Fortunately, there has been tremendous progress bridging information processing
and the nonequilibrium thermodynamics required to support it
\cite{Jarz11aa,Parr15aa}. This information thermodynamics addresses processes
that range from the very small scale, such as the operation nanoscale devices
and molecular dynamics \cite{Saga12a}, to the cosmologically large, such the
character and evolution of black holes \cite{Fiol94aa,Das04aa}. Recent
technological innovations allowed many of the theoretical advances to
be experimentally verified \cite{Beru12, Toya10a}. The current state of
knowledge in this rapidly evolving arena is reviewed in Refs.
\cite{Maru2009,Seki10aa,Seif12aa}. Here, we use information thermodynamics to
describe the physical limits on random number generation. Though the latter is
often only treated as a purely abstract mathematical subject, practicing
scientists and engineers know how essential random number generation is in
their daily work. The following explores the underlying necessary thermodynamic
resources.

First, Sec. \ref{sec:RNG} addresses random number generation, analyzing the
thermodynamics of three algorithms, and discusses physical implementations.
Second, removing the requirement of an input randomness source, Sec.
\ref{sec:PRNG} turns to analyze pseudorandom number generation and its costs.
Third, Sec.  \ref{sec:TRNG} analyzes the thermodynamics of true random number
generation. Finally, the conclusion compares the RNG strategies and their costs
and suggests future problems.

\section{Random Number Generation} 
\label{sec:RNG} 

Take a fair coin as our source of randomness.\footnote{Experiments reveal this
assumption is difficult if not impossible to satisfy. Worse, if one takes the
full dynamics into account, a flipped physical coin is quite predictable
\cite{Diac07a}.} Each flip results in a Head or a Tail with $50\%-50\%$
probabilities. However, we need a coin that $1/4$ of the time generates Heads
and $3/4$ of the time Tails. Can the series of fair coin flips be transformed?
One strategy is to flip the coin twice. If the result is Head-Head, we report
Heads. Else, we report Tails. The reported sequence is equivalent to flipping a
coin with a bias $1/4$ for Heads and $3/4$ for Tails.

Each time we ask for a sample from the biased distribution we must flip the
fair coin twice. Can we do better? The answer is yes. If the first flip results
in a Tail, independent of the second flip's result, we should report Tail. We
can take advantage of this by slightly modifying the original strategy. If the
first flip results in a Tail, stop. Do not flip a second time, simply report a
Tail, and start over. With this modification, $1/2$ of the time we need a
single flip and $1/2$ the time we need two flips. And so, on average we need
$1.5$ flips to generate the distribution of interest. This strategy reduces the
use of the fair coin ``resource'' by 25\%.

Let's generalize. Assume we have access to a source of randomness that
generates the distribution $\{p_i: i \in \ProcessAlphabet\}$ over discrete
alphabet $\ProcessAlphabet$. We want an algorithm that generates another target
distribution $\{q_j: j \in \mathcal{B}\}$ from samples of the given source. (Generally, the source
of randomness $\{p_i\}$ can be known or unknown to us.) In this, we ask for a
single correct sample from the target distribution. This is the \emph{immediate
random number generation problem}: Find an algorithm that minimizes the
expected number of necessary samples of the given source to generate one sample
of the target.\footnote{A companion is the \emph{batch random number generation
problem}: Instead of a single sample, generate a large number of inputs and
outputs. The challenge is to find an algorithm minimizing the ratio of the
number of inputs to outputs \cite{Elia72a,Pere92a,Romi99a}.}

The goal in the following is to analyze the thermodynamic costs when these
algorithmically efficient algorithms are implemented in a physical substrate.
This question parallels that posed by Landauer \cite{Land61a,Benn82}: What is
the minimum thermodynamic cost to erase a bit of information? That is, rather
than destroying information, we analyze the costs of creating information with
desired statistical properties given a source of randomness.

\begin{figure}
\includegraphics[width=1\linewidth]{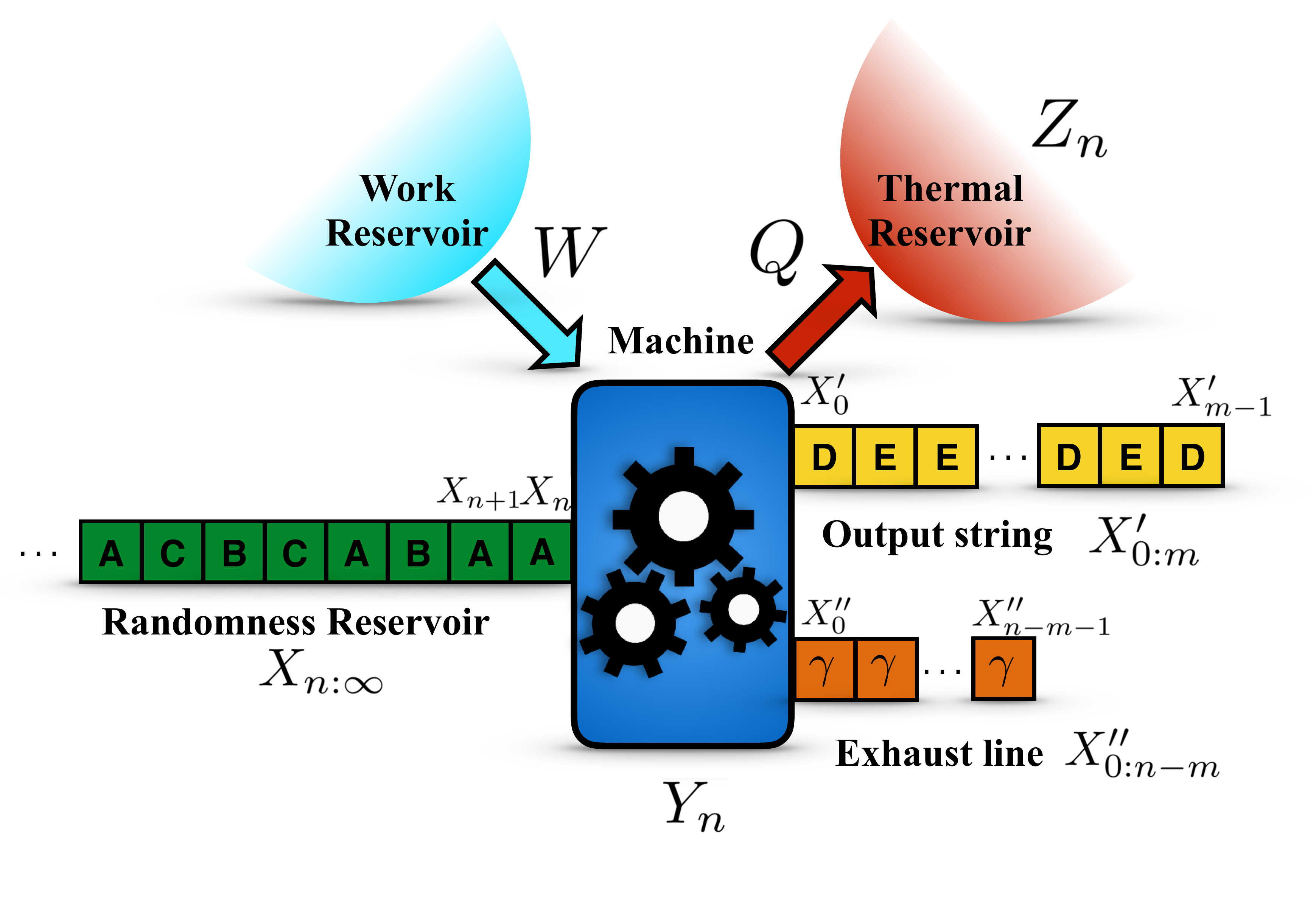}
\caption{Thermodynamically embedded finite-state machine implementing an
	algorithm that, from the source of randomness available on the input
	string, generates random numbers on the output string obeying a desired
	target distribution and an exhaust with zero entropy. Input string and
	output string symbols can come from different alphabet sets. For example,
	here the input symbols come from the set $\{A,B,C\}$ and the outputs from
	$\{D,E\}$.  Exhaust line symbols all are the same symbols $\gamma$.
	}
\label{fig:PhysicalRNG}
\end{figure}

\paragraph*{Bounding the Energetics:} 
The machine implementing the algorithm transforms symbols on an input string
sampled from an information reservoir to an output symbol string and an exhaust
string, using a finite-state machine that interacts with heat and work
reservoirs; see Fig.~\ref{fig:PhysicalRNG}. The input \emph{Randomness
Reservoir} is the given, specified source of randomness available to the RNG.
The states and transition structure of the finite-state machine implement the
RNG algorithm. The output string is then the samples of distribution of
interest. The exhaust string is included to preserve state space.

%\alert{What about the energetics of the ratchet? We need to assign energy
%levels to its states. Or do we? Do those levels lead to detailed-balance and
%so microscopically reversible physical dynamics? What about non-detailed
%balanced energetics and irreversible microscopic dynamics? Does this
%energetics change or parametrize the bounds? Without this, we are really not
%determining the heat properly (the bounds' LHS), instead just determining (in
%substantial detail) the information processing (the bounds' RHS).}

Here, we assume inputs $X_n$ are independent, identically distributed
(IID) samples from the randomness reservoir with
discrete alphabet $\ProcessAlphabet$. The output includes two strings, one with
samples from the target distribution $X^{\prime}_{m}$ over alphabet
$\mathcal{B}$ and another, the exhaust string. At each step one symbol,
associated with variable $X_n$, enters the machine. After analyzing that symbol
and, depending on its value and that of previous input symbols, the machine
either writes a symbol to the output string or to the exhaust string. $Y_n$
denotes the machine's state at step $n$ after reading input symbol $X_n$.
The last symbol in the output string after the input $X_n$ is read is
denoted $X^{\prime}_m$, where $m \leq n$ is not necessarily equal to $n$. The
last symbol in the exhaust string is $X^{\prime\prime}_{n-m}$. As a result, the
number of input symbols read by the machine equals the number of symbols
written to either the output string or the exhaust string. To guarantee that
the exhaust makes no thermodynamic contribution, all symbols written to
$X^{\prime\prime}_{i}$s are the same---denoted $\gamma$. Without loss of
generality we assume both the input and output sample space is
$\ProcessAlphabet \cup \mathcal{B} \cup \{\gamma\}$. In the following we refer
to the random-variable input chain as $X_{n:\infty} = X_{n}X_{n+1} \cdots
X_{\infty}$, output chain as $X^{\prime}_{0:m} = X^{\prime}_{0}X^{\prime}_{1}
\cdots X^{\prime}_{m-1}$, and exhaust chain as $X^{\prime\prime}_{0:n-m} =
X^{\prime\prime}_{0}X^{\prime\prime}_{1} \cdots X^{\prime\prime}_{n-m-1}$.

The machine also interacts with an environment consisting of a \emph{Thermal
Reservoir} at temperature $T$ and a \emph{Work Reservoir}. The thermal
reservoir is that part of the environment which contributes or absorbs heat,
exchanging thermodynamic entropy and changing its state $Z_n$. The work
reservoir is that part which contributes or absorbs energy by changing its
state, but without an exchange of entropy. All transformations are performed
isothermally at temperature $T$. As in Fig.~\ref{fig:PhysicalRNG}, we denote
heat that flows to the thermal reservoir by $Q$. To emphasize, $Q$ is positive
if heat flows into the thermal reservoir. Similarly, $W$ denotes the work
done on the machine and not the work done by the machine.\footnote{Several
recent works \cite{Mand012a,Bara2013,Boyd15a} use the same convention for $Q$,
but $W$ is defined as the work done by the machine. This makes sense in
those settings, since the machine is intended to do work.}

After $n$ steps the machine has read $n$ input symbols and generated $m$ output
symbols and $n-m$ exhaust symbols. The thermodynamic entropy change of the entire system is \cite[App.
A]{Boyd15a}:
\begin{align*}
\Delta S \equiv \kB \ln 2
  \big(&
  \H[X^{\prime\prime}_{0:n-m},X^{\prime}_{0:m},X_{n:\infty},Y_n,Z_n] \\- &\H[X_{0:\infty},Y_0,Z_0]
  \big)
  ~,
\end{align*}
where $\H[\cdot]$ is the Shannon entropy \cite{Cove06a}. Recalling the
definition of mutual information $\I[\cdot:\cdot]$ \cite{Cove06a}, we rewrite
the change in Shannon entropy on the righthand side as:
\begin{align*}
\Delta \H &= (\H[X^{\prime\prime}_{0:n-m},X^{\prime}_{0:m},X_{n:\infty},Y_n] - \H[X_{0:\infty},Y_0]) \\
& + (\H[Z_n] - \H[Z_0]) \\
& - (\I[X^{\prime\prime}_{0:n-m},X^{\prime}_{0:m},X_{n:\infty},Y_n:Z_n] - \I[X_{0:\infty},Y_0:Z_0])
~.
\end{align*}

By definition, a heat bath is not correlated with other subsystems, in
particular, with portions of the environment. As a result, both mutual
informations vanish. The term $\H[Z_n] - \H[Z_0]$ is the heat bath's entropy
change, which can be written in terms of the dissipated heat $Q$:
\begin{align*}
\H[Z_n] - \H[Z_0] = \frac{Q}{\kB T \ln 2}
  ~.
\end{align*}
Since by assumption the entire system is closed, the Second Law of
Thermodynamics says that $\Delta S \geq 0$. Using these relations gives:
\begin{align*}
Q \geq -\kB T \ln 2
  \big( \H[X^{\prime\prime}_{0:n-m},X^{\prime}_{0:m},X_{n:\infty},Y_n] - \H[X_{0:\infty},Y_0] \big)
  ~.
\end{align*}
To use rates we divide both sides by $n$ and decompose the first joint entropy:
\begin{align*}
\frac{Q}{n} \geq &-\frac{\kB T \ln 2}{n}
  \big( \H[X^{\prime\prime}_{0:n-m},X^{\prime}_{0:m},X_{n:\infty}]
  - \H[X_{0:\infty}] \\
  & \quad + \H[Y_n] - \H[Y_0]
    - \I[X^{\prime\prime}_{0:n-m},X^{\prime}_{0:m},X_{n:\infty}:Y_n] \\
  & \quad + \I[X_{0:\infty}:Y_0] \big)   
    ~.
\end{align*}

Appealing to basic information identities, a number of the righthand terms
vanish, simplifying the overall bound. First, since the Shannon entropy of a
random variable $Y$ is bounded by logarithm of the size $|\MeasAlphabet_Y|$
of its state space, we have for the ratchet's states:
\begin{align*}
\lim_{n \to \infty} \frac{1}{n} \H[Y_n]
  & = \lim_{n \to \infty} \frac{1}{n} \H[Y_0] \\
  & \leq \lim_{n \to \infty} \frac{1}{n} \log_2|\MeasAlphabet_Y| \\
  & = 0~,
\end{align*}
Second, recalling that the two-variable mutual information is nonnegative and
bounded above by the Shannon entropy of the individual random variables, in the
limit $n \to \infty$ we can write:
\begin{align*}
\lim_{n \to \infty} \frac{1}{n}\I[X^{\prime\prime}_{0:n-m},X^{\prime}_{0:m},X_{n:\infty}:Y_n] 
  & \leq \lim_{n \to \infty} \frac{1}{n} \H[Y_0] \\
  & = 0
  ~.
\end{align*}
Similarly, $\lim_{n \to \infty} \frac{1}{n} \I[X_{0:\infty}:Y_0]=0$.
As a result, we have:
\begin{align*}
\frac{Q}{n} & \geq -\frac{\kB T \ln 2}{n}
  \big( \H[X^{\prime\prime}_{0:n-m},X^{\prime}_{0:m},X_{n:\infty}]
  - \H[X_{0:\infty}] \big)
    ~.
\end{align*}
%\begin{align*}
%\frac{Q}{n} & \geq -\frac{\kB T \ln 2}{n}
%  \big( \H[X^{\prime\prime}_{0:n-m},X^{\prime}_{0:m},X_{n:\infty}] - \H[X_{0:\infty}] \\
%  & \qquad + \H[Y_n] - \H[Y_0]
%    - \I[X^{\prime\prime}_{0:n-m},X^{\prime}_{0:m},X_{n:\infty}:Y_n] + \I[X_{0:\infty}:Y_0] \big) \\
%  & = -\frac{\kB T \ln 2}{n}
%      \big( \H[X^{\prime}_{0:m}] + \H[X_{n:\infty}] - \H[X_{0:\infty}] \\
%  & \qquad + \H[Y_n] - \H[Y_0] - \I[X^{\prime}_{0:m}:X_{n:\infty}] \\
%  & \qquad - \I[X^{\prime}_{0:m},X_{n:\infty}:Y_n] + \I[X_{0:\infty}:Y_0]
%  \big)
%  ~.
%\end{align*}
We can also rewrite the joint entropy as:
\begin{align*}
\H[X^{\prime\prime}_{0:n-m},X^{\prime}_{0:m},X_{n:\infty}]  &= 
\H[X^{\prime}_{0:m},X_{n:\infty}] + \H[X^{\prime\prime}_{0:n-m}] \\
& \quad - \I[X^{\prime}_{0:m},X_{n:\infty}:X^{\prime\prime}_{0:n-m}]
  ~.
\end{align*}
Since the entropy of exhaust vanishes, $\H[X^{\prime\prime}_{0:n-m}]=0$.
Also, $\I[X^{\prime}_{0:m},X_{n:\infty}:X^{\prime\prime}_{0:n-m}]$ is bounded
above by it, $\I[X^{\prime}_{0:m},X_{n:\infty}:X^{\prime\prime}_{0:n-m}]$
also vanishes. This leads to:
\begin{align*}
\H[X^{\prime\prime}_{0:n-m},X^{\prime}_{0:m},X_{n:\infty}]  &= 
\H[X^{\prime}_{0:m},X_{n:\infty}] 
  ~.
\end{align*}
This simplifies the lower bound on the heat to: 
\begin{align*}
\frac{Q}{n} & \geq -\frac{\kB T \ln 2}{n}
  \big( \H[X^{\prime}_{0:m},X_{n:\infty}] - \H[X_{0:\infty}] \big)
    ~.
\end{align*}
Rewriting the righthand terms, we have:
\begin{align*}
\H[X_{0:\infty}] = \H[X_{0:n}] + \H[X_{n:\infty}] - \I[X_{0:n}:X_{n:\infty}]
\end{align*}
and
\begin{align*}
\H[X^{\prime}_{0:m},X_{n:\infty}] = \H[X^{\prime}_{0:m}] + \H[X_{n:\infty}] - \I[X^{\prime}_{0:m}:X_{n:\infty}]
  ~.
\end{align*}
These lead to:
\begin{align*}
\frac{Q}{n} & \geq -\frac{\kB T \ln 2}{n} \big( \H[X^{\prime}_{0:m}] - \H[X_{0:n}]\\
  & \qquad + \I[X_{0:n}:X_{n:\infty}] - \I[X^{\prime}_{0:m}:X_{n:\infty}]
  \big)
  ~.
\end{align*}

Since the inputs are IID, $\I[X_{0:n}:X_{n:\infty}]$ vanishes. Finally,
$\I[X^{\prime}_{0:m}:X_{n:\infty}]$ is bounded above by
$\I[X_{0:n}:X_{n:\infty}]$, meaning that $\I[X^{\prime}_{0:m}:X_{n:\infty}]=0$.
Using these we have:
\begin{align*}
\frac{Q}{n} \geq \frac{\kB T \ln 2}{n}
  \big( \H[X_{0:n}] - \H[X^{\prime}_{0:m}] \big)
  ~.
\end{align*}
This can be written as:
\begin{align*}
\frac{Q}{n} \geq \kB T \ln 2
  \left(
  \frac{\H[X_{0:n}]}{n}
  -\frac{\H[X^{\prime}_{0:m}]}{m} \left( \frac{m}{n} \right)
  \right)
  ~.
\end{align*}

As $n \to \infty$, $\H[X_{0:n}]/n$ converges to the randomness reservoir's Shannon
entropy rate $h$ and $\H[X^{\prime}_{0:m}]/m$ converges to the output's entropy
rate $h^\prime$. The tapes' relative velocity term $m/n$ also converges and we
denote the limit as $1/\Lhat$. As a result, we have the rate $\dot{Q}$ of heat
flow from the RNG machine to the heat bath:
\begin{align}
\dot{Q} \geq \kB T \ln 2 \left( h - \frac{h^{\prime}}{\Lhat} \right)
  ~.
\label{QBound}
\end{align}

Since the machine is finite state, its energy is bounded. In turn, this means
the average energy entering the machine, above and beyond the constant amount
that can be stored, is dissipated as heat. In other words, the average work
rate $\dot{W}$ and average heat dissipation rate $\dot{Q}$ per input are equal:
$\dot{W} = \dot{Q}$.

\newcommand{\QLB} { Q_\text{LB} } 

This already says something interesting. To generate one random number the
average change $\Delta W$ in work done on the machine and the average change
$\Delta Q$ in heat dissipation by the machine are directly related: $\Delta W =
\Delta Q = \Lhat \dot{Q}$. More to the point, denoting the lower bound by $\QLB
\equiv \kB T \ln 2 \left( \Lhat h - h^{\prime} \right)$ immediately leads to a
Second Law adapted to RNG thermodynamics:
\begin{align}
\Delta Q \geq \QLB  ~.
\label{DQBound}
\end{align}
It can be shown that $\Lhat$ is always larger or equal to $h^\prime/h$
\cite{Cove06a} and so $\QLB \geq 0$.\footnote{This is not generally true for
the setup shown in Fig.~ \ref{fig:PhysicalRNG} interpreted most broadly. For
computational tasks more general than RNG, $Q_{LB}$ need not be positive.} This
tells us that RNG algorithms are always \emph{heat dissipative} or, in other
words, \emph{work consuming} processes. Random numbers generated by RNGs cost
energy. This new RNG Second Law allows the machine to take whatever time it
needs to respond to and process an input. The generalization moves the
information ratchet architecture \cite{Boyd15a} one step closer to that of
general Turing machines \cite{Lewi98a}, which also take arbitrary time to
produce an output. We now apply this generalized Second Law to various
physically embedded RNG algorithms.

\paragraph*{von Neumann RNG:} Consider the case where the randomness resource is
a biased coin with \emph{unknown} probability $p \neq 1/2$ for Heads. How can we use
this imperfect source to generate fair (unbiased $p = 1/2$) coin tosses using
the minimum number of samples from the input? This
problem was first posed by von Neumann \cite{Vneu51}. The answer is
simple but clever. What we need is a symmetry to undo the source's bias
asymmetry. The strategy is to flip the biased coin twice. If the result
is Heads-Tails we report a Head; if it is Tails-Heads we report Tails.
If it is one of the two other cases, we neglect the flips and simply repeat
from the beginning. A moment's reflection reveals that using any source of
randomness that generates independent, identically distributed (IID) samples
can be used in this way to produce a statistically uniform sample, even if we
do not know the source's bias.

Note that we must flip the biased coin more than twice, perhaps many more, to
generate an output. More troublesome, there is no bound on how many times we
must flip to get a useful output.

\begin{figure}
\includegraphics[width=1\linewidth]{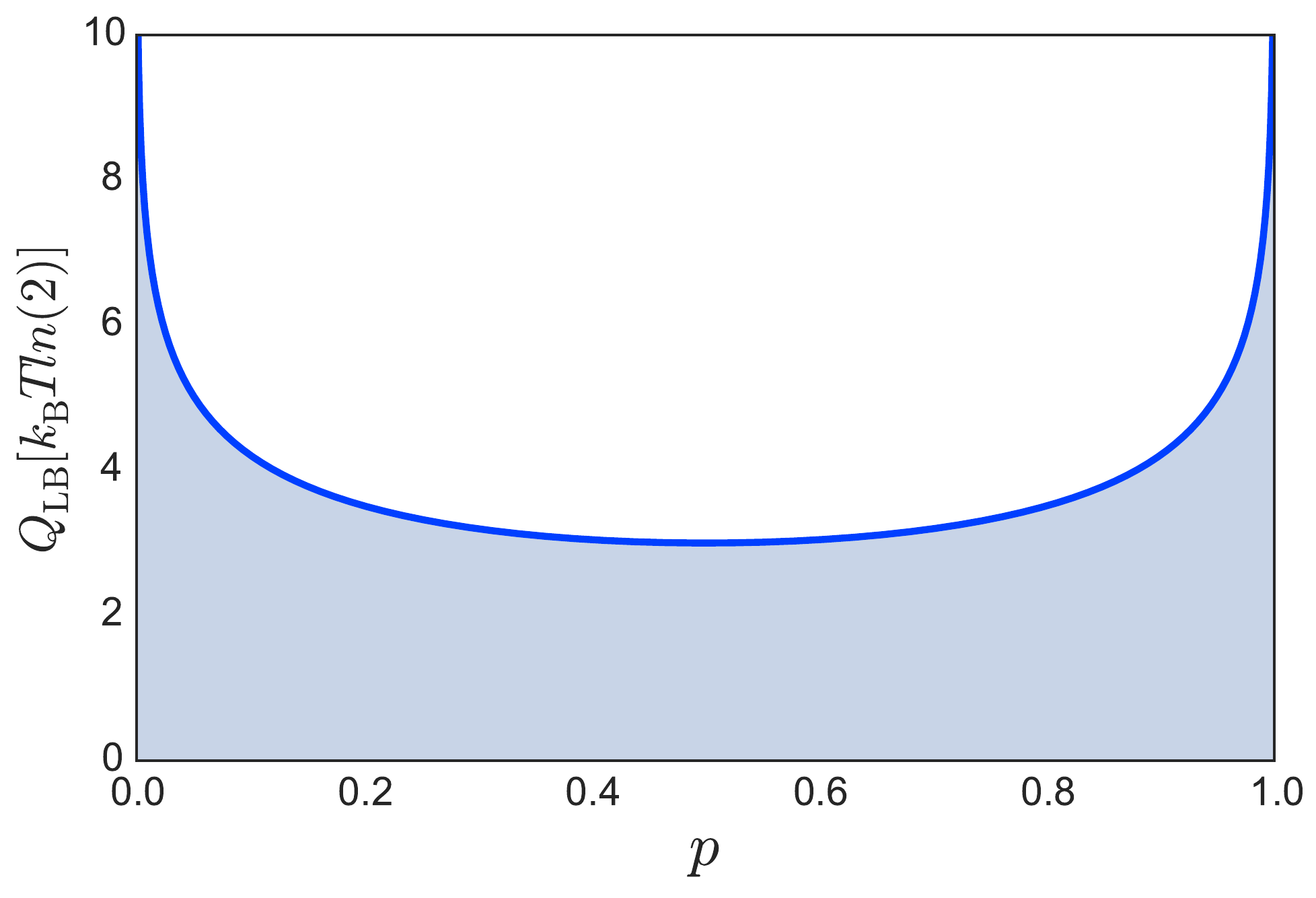}
\caption{Lower bound on heat dissipation during the process of single fair
	sample generation by von Neumann algorithm versus the input bias $p$.
	}
\label{VNBOUND}
\end{figure}

So, what are the thermodynamic costs of this RNG scheme? With probability
$2p(1-p)$ the first two flips lead to an output; with probability
$(1-2p(1-p))(2p(1-p))$ the two flips do not, but the next two flips will; and
so on. The expected number of flips to generate a fair coin output is
$\Lhat = \frac{1}{p(1-p)}$. Using Eq.~(\ref{DQBound}) this costs:
\begin{align}
\label{VNB}
\QLB = \kB T \ln 2
  \left( \frac{\H(p)}{p(1-p)} -1 \right)
~.
\end{align}
Figure~\ref{VNBOUND} shows $\QLB$ versus source bias $p$. 
It is always positive with a minimum $3\kB T \ln 2$ at $p=\sfrac{1}{2}$. 

This minimum means that generating a fair coin from a fair coin has a heat cost
of $3\kB T \ln 2$. At first glance, this seems wrong. Simply pass the fair coin
through. The reason it is correct is that the von Neumann RNG does not know the
input bias and, in particular, that it is fair. In turn, this means we may flip
the coin many times, depending on the result of the flips, costing energy.

Notably, the bound diverges as $p \to 0$ and as $p \to 1$, since the RNG must
flip an increasingly large number of times. As with all RNG methods, the
positive lower bound implies that generating an unbiased sample via the von
Neumann method is a \emph{heat dissipative} process. We must put energy in to
get randomness out.

Consider the \emph{randomness extractor} \cite{Trev00a}, a variation on von
Neumann RNG at extreme $p$, that uses a weakly random physical source but still
generates a highly random output. (Examples of weakly random sources include
radioactive decay, thermal noise, shot noise, radio noise, avalanche noise in
Zener diodes, and the like. We return to physical randomness sources shortly.)
For a weakly random source $p \ll 1$, the bound in Eq.~(\ref{VNB}) simplifies
to $- \kB T \ln p$, which means heat dissipation diverges at least as fast as $-
\ln p$ in the limit $p \to 0$.

\paragraph*{Knuth and Yao RNG:}  
Consider a scenario opposite von Neumann's where we have a fair coin and can
flip it an unlimited number of times. How can we use it to generate samples
from any desired distribution over a finite alphabet using the minimum number
of samples from the input? Knuth and Yao were among the first to attempt an
answer \cite{Knut76a}. They proposed the \emph{discrete distribution generation
tree} (DDD-tree) algorithm.

The algorithm operates as follows. Say the target distribution is $\{p_j\}$
with probabilities $p_j$ ordered from large to small. Define the partial sum
$\beta_k = \sum_{j=1}^{k} p_j$, with $\beta_0=0$. This partitions
the unit interval $(0,1)$ into the subintervals $(\beta_{k-1},\beta_k)$ with
lengths $p_k$. Now, start flipping the coin, denoting the outcomes $X_1, X_2,
\ldots$. Let $S_l = \sum_{m=1}^{l} X_m 2^{-m}$. It can be easily shown that
$S_{\infty}$ has the uniform distribution over the unit interval. At any step
$l$, when we flip the coin, we examine $S_l$. If there exists a $k$ such
that:
\begin{align}
\label{eq:RelKY}
\beta_{k-1} \leq S_l<S_l + 2^{-l} \leq \beta_{k}
~,
\end{align}
the output generated is symbol $k$. If not, we flip the coin again for 
$l+1$ or more times until we find a $k$ that satisfies the relation in
Eq.~(\ref{eq:RelKY}) and report that $k$ as the output.

This turns on realizing that if the condition is satisfied, then the value
of future flips does not matter since, for $r>l$, $S_r$ always falls in the
subinterval $(\beta_{k-1},\beta_k)$. Recalling that $S_{\infty}$ is uniformly
distributed over $(0,1)$ establishes that the algorithm generates the desired
distribution $\{p_j\}$. The algorithm can be also interpreted as walking a
binary tree,\footnote{For details see Ref. \cite{Cove06a}.} a view related to
arithmetic coding \cite{Cove06a}. Noting that the input has entropy rate $h=1$
and using Eq.~(\ref{QBound}) the heat dissipation is bounded by:
\begin{align}
\QLB = \kB T \ln 2 \left( \Lhat - \H[\{p_i\}] \right)~.
\label{QKY}
\end{align}

Now, let's determine $\Lhat$ for the Knuth-Yao RNG. Ref. \cite{Knut76a} showed
that:
\begin{align}
\H[\{p_i\}] \leq \Lhat \leq \H[\{p_i\}] + 2
  ~.
\label{KYL}
\end{align}
More modern proofs are found in Refs. \cite{Romi99a} and \cite{Cove06a}.
Generally, given a general target distribution the Knuth-Yao RNG's $\Lhat$ can
be estimated more accurately. However, it cannot be calculated in closed form,
only bounded. Notably, there are distributions $\{p_j\}$ for which $\Lhat$ can
be calculated exactly. These include the \emph{dyadic distributions} whose
probabilities can be written as $2^{-n}$ with $n$ an integer. For these target
distributions, the DDG-tree RNG has $\Lhat = \H[\{p_i\}]$.

Equations~(\ref{DQBound}) and (\ref{KYL}) lead one to conclude that the heat
dissipation for generating one random sample is always a strictly positive
quantity, except for the dyadic distributions which lead to vanishing or
positive dissipation. Embedding the DDG-tree RNG into a physical machine, this
means one must inject work to generate a random sample. The actual amount of
work depends on the target distribution given.

\begin{table} 
\centering
\begin{tabular}{|l c|} 
 \hline
Input & Output\\ [0.5ex] 
 \hline\hline
 $00$ & $A$  \\ 
 \hline
$01$ & $B$  \\ 
 \hline
$10$ & $C$ \\ 
 \hline
$110$ & $B$ \\
 \hline
 $1110$ & $A$ \\
 \hline
 $11110$ & $A$ \\
 \hline
 $111110$ & $B$ \\
 \hline
 $111111$ & $C$ \\
 \hline
\end{tabular}
\caption{Most efficient map from inputs to outputs when using the DDG-tree
	RNG method.
	}
\label{tab:TableIO}
\end{table}

Let us look at a particular example. Consider the case that our source of
randomness is a fair coin with half and half probability over symbols $0$ and
$1$ and we want to generate the target distribution $\{ \sfrac{11}{32},
\sfrac{25}{64}, \sfrac{17}{64} \}$ over symbols $A, B$, and $C$. The target
distribution has Shannon entropy $\H[\{p_i\}] \approx 1.567$ bits.
Equation~(\ref{KYL}) tells us that $\Lhat$ should be larger than this. The
DDG-tree method leads to the most efficient RNG. Table \ref{tab:TableIO} gives
the mapping from binary inputs to three-symbol outputs. $\Lhat$ can be
calculated using the table: $\Lhat \approx 2.469$. This is approximately $1$
bit larger than the entropy consistent with Eq.~(\ref{KYL}). Now, using
Eq.~(\ref{QKY}), we can bound the dissipated heat: $\QLB \approx 0.625 \kB T$.

\paragraph*{Roche and Hoshi RNG:} 
A more sophisticated and more general RNG problem was posed by Roche in 1991
\cite{Roch91a}: What if we have a so-called $M$-coin that generates the
distribution $\{p_i: i = 1, \ldots, M\}$ and we want to use it to generate a
different target distribution $\{q_j\}$? Roche's algorithm was probabilistic.
And so, since we assume the only source of randomness to which we have access
is the input samples themselves, Roche's approach will not be discussed here.

However, in 1995 Hoshi introduced a deterministic algorithm \cite{Hosh97a}
from which we can determine the thermodynamic cost of this general RNG
problem. Assume the $p_i$s and $q_j$s are ordered from large to small. Define
$\alpha_t = \sum_{i=1}^{t} p_i$ and $\beta_k = \sum_{j=1}^{k} q_j$, with
$\alpha_0=\beta_0=0$. These quantities partition $(0,1)$ into subintervals
$[\alpha_{t-1},\alpha_t)$ and $B_k\equiv[\beta_{k-1},\beta_k)$ with lengths
$p_t$ and $q_k$, respectively. Consider now the operator $\mathcal{D}$ that
takes two arguments---an interval and an integer---and outputs
another interval:
\begin{align*}
\mathcal{D} ([a,b) , t) = [ a + (b-a)\alpha_{t-1}, a + (b-a)\alpha_{t})
~.
\end{align*}

Hoshi's algorithm works as follows. Set $n=0$ and $R_0 = [0,1)$. Flip the
$M$-coin, call the result $x_n$. Increase $n$ by one and set $R_n =
\mathcal{D}(R_{n-1} , x_n)$. If there is a $k$ such that $R_n \subseteq B_k$,
then report $k$, else flip the $M$-coin again.

Han and Hoshi showed that  \cite{Hosh97a}:
\begin{align*}
\frac{\H[\{q_j\}]}{\H[\{p_i\}]}
  \leq \Lhat \leq \frac{\H[\{q_j\}] + f(\{p_i\})}{\H[\{p_i\}]}
~,
\end{align*}
where:
\begin{align*}
f(\{p_i\}) = \ln(2(M-1))
  + \frac{\H[\{p_{max}, 1 - p_{max}\}]}{1 - p_{max}}
~,
\end{align*}
with $p_{max} = \max \limits_{i=1,\cdots,M} p_i$. Using this and
Eq.~(\ref{DQBound}) we see that the heat dissipation per sample is always
positive except for measure-zero cases for which the dissipation may be zero or
not. This means one must do work on the system independent of input and output
distributions to generate the target sample. Again, using this result and
Eq.~(\ref{DQBound}) there exist input and output distributions with heat dissipation at least as large as $\kB T \ln 2f(\{p_i\})$.

\paragraph*{RNG Physical Implementations:} 
Recall the first RNG we described. The input distribution is a fair coin and
the output target distribution is a biased coin with bias $\sfrac{1}{4}$. Table
\ref{TableIPE} summarizes the optimal algorithm. Generally, optimal algorithms
require the input length to differ from the output length---larger than or equal, respectively.

This is the main challenge to designing physical implementations. Note that for
some inputs, after they are read, the machine should wait for additional inputs
until it receives the correct input and then transfers it deterministically to
the output. For example, in our problem if input $0$ is read, the output would
be $0$. However, if $1$ is read, the machine should wait for the next input
and then generate an output. How to implement these delays? Let's explore a
chemical implementation of the algorithm.

\begin{table} 
\centering
\begin{tabular}{|c c|} 
 \hline
Input & Output\\ [0.5ex] 
 \hline\hline
 $0$ & $0$  \\ 
 \hline
$10$ & $0$  \\ 
 \hline
 $11$ & $1$  \\ 
 \hline
\end{tabular}
\caption{Immediate random number generation: The most efficient map from inputs
	to output to transform fair coin inputs to biased coin outputs with bias
	$\sfrac{1}{4}$.
	}
\label{TableIPE}
\end{table}

\emph{Chemical reaction networks} (CRNs) \cite{Temk96a,Cook09a} have been
widely considered as substrates for physical information processing
\cite{Jian12a} and as a programming model for engineering artificial systems
\cite{Magn97a,Hjel91a}. Moreover, CRN chemical implementations have been
studied in detail \cite{Card11a,Solo10a}. CRNs are also efficiently
Turing-universal \cite{Solo08a}, which power makes them appealing. One of their
main applications is deterministic function computation
\cite{Chen14a,Doty15aa}, which is what our RNGs need.

\begin{figure}
\includegraphics[width=0.7\linewidth]{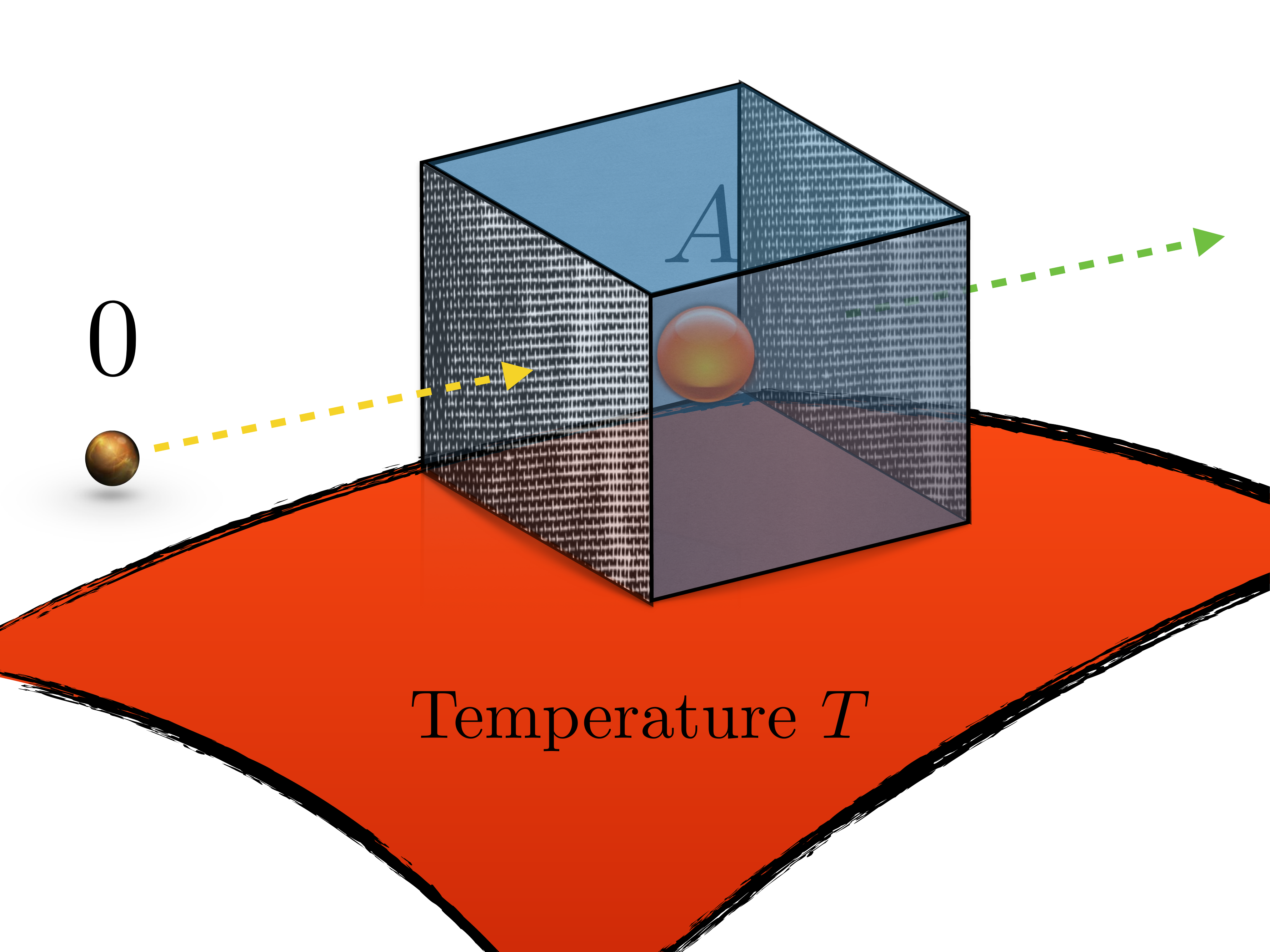}
\caption{Chemical Reaction Network (CRN) implementation of an RNG machine
	consisting of a box and a particle in it. The left wall acts as a membrane
	filter such that only input particles, $0$ and $1$, can enter, but no
	particles can exit through the wall. The right wall is also a membrane
	designed such that only output particles, $0$, $1$ and $\gamma$, can exit. At the
	beginning the only particle in the box is ``machine particle'' $A$, which is
	confined to stay in the box. Every $\tau$ seconds a new input particle
	enters the box from the left and, depending on the reaction between the
	input particle and the machine particle, an output particle may or may not
	be generated that exists through the right wall.
	}
\label{PHYSEXAM}
\end{figure}

Consider five particle types---$0$, $1$, $A$, $B$, and $\gamma$---and a machine
consisting of a box that can contain them. Particles $0$ and $1$ can be inputs
to or outputs from the machine and particle $\gamma$ can be an output from the
machine. ``Machine'' particles $A$ and $B$ always stay in the machine's box and
are in contact with a thermal reservoir. Figure~\ref{PHYSEXAM} shows that the
left wall is designed so that only input particles ($0$ and $1$) can enter, but
no particles can exit. The right wall is designed so that only output particles
($0$, $1$, and $\gamma$) can exit.

To get started, assume there is only a single machine particle $A$ in the box.
Every $\tau$ seconds a new input particle, $0$ or $1$, enters from the left.
Now, the particles react in the following way:
\begin{align*}
& 0 + A\ \Rightarrow\ A + 0 ~,\\
& 1 + A\ \Rightarrow\ B ~,\\
& 0 + B\ \Rightarrow\ A + 0 + \gamma~,\\
& 1 + B\ \Rightarrow\ A + 1 + \gamma
  ~.
\end{align*}
The time period of each chemical reaction is also $\tau$. With this assumption
it is not hard to show that if the distribution of input particles $0$ and $1$
is $\{\sfrac{1}{2}, \sfrac{1}{2} \}$ then the distribution of output particles
$0$ and $1$ would be $\{\sfrac{3}{4}, \sfrac{1}{4} \}$, respectively. Thus,
this CRN gives a physical implementation of our original RNG.

Using Eq.~(\ref{DQBound}) we can put a lower bound on the average heat
dissipation per output: $\QLB \approx 0.478 \kB T$. Since deriving the bound
does not invoke any constraints over input or output particles, the bound is a
universal lower bound over all possible reaction energetics. That is, if we
find any four particles (molecules) obeying the four reactions above then the
bound holds. Naturally, depending on the reactions' energetics, the CRN-RNG's
$\Delta Q$ can be close to or far from the bound. Since CRNs are
Turing-universal \cite{Solo08a} they can implement all of the RNGs studied up
to this point. The details of designing CRNs for a given RNG algorithm can be
gleaned from the general procedures given in Ref. \cite{Chen14a}.

\section{Pseudorandom Number Generation} 
\label{sec:PRNG} 

So far, we abstained from von Neumann's sin by assuming a source of
randomness---a fair coin, a biased coin, or any general IID process.
Nevertheless, modern digital computers generate random numbers using purely
deterministic arithmetical methods. This is \emph{pseudorandom number
generation} (PRNG). Can these methods be implemented by finite-state machines?
Most certainly. The effective memory in these machines is very large, with the
algorithms typically allowing the user to specify the amount of state
information used \cite{BSD5a}. Indeed, they encourage the use of large amounts
of state information, promising better quality random numbers in the sense
that the recurrence time (generator period) is astronomically large. Our
concern, though, is not analyzing their implementations. See Ref.
\cite{Knut81a} for a discussion of design methods. We can simply assume they
can be implemented or, at least, there exist ones that have been, such as the
Unix C-library \texttt{random()} function just cited.

The PRNG setting forces us to forego accessing a source of randomness. The
input randomness reservoir is not random at all. Rather, it is simply a pulse
that indicates that an output should be generated. Thus, $h = 0$ and $\Lhat
=1$. In our analysis, we can take the outputs to be samples of any
desired IID process.

Even though a PRNG is supposed to generate a random number, in reality after
setting the seed \cite{Gent13,Rubi98a} it, in fact, generates an exactly
periodic sequence of outputs. Thus, as just noted, to be a good PRNG algorithm
that period should be relatively long compared to the sample size of interest.
Also, the sample statistics should be close to those of the desired
distribution. This means that if we estimate $h^\prime$ from the sample it
should be close to the Shannon entropy rate of the target distribution.
However, in reality $h^\prime=0$ since $h^\prime$ is a measure over
infinite-length samples, which in this case are completely nonrandom due to
their periodicity.

This is a key point. When we use PRNGs we are only concerned about samples with
comparatively short lengths compared to the PRNG period. However, when
determining PRNG thermodynamics we average over asymptotically large samples.
As a result, we have $\QLB = 0$ or, equivalently, $\Delta Q \geq 0$. And so,
PRNGs are potentially heat dissipative processes. Depending on the PRNG
algorithm, it may be possible to find machinery that achieves the lower bound
(zero) or not. To date, no such PRNG implementations have been introduced.

Indeed, the relevant energetic cost bounds are dominated by the number of
logically irreversible computation steps in the PRNG algorithm, following
Landauer \cite{Land61a}. This, from a perusal of open source code for modern
PRNGs, is quite high. However, this takes us far afield, given our focus on
input-output thermodynamic processing costs.

%\begin{figure}
%\includegraphics[width=0.8\linewidth]{Image/DET3.pdf}
%\caption{True biased-coin generator emits random bits with bias $p$. The
%	machine has one internal state $S$ and inputs and outputs can be
%	either $1$ or $0$. All the states have zero energy. The system is designed
%	such that the joint states $0\otimes S$ and $1\otimes S$ have 
%	energies $0$ and $\Delta E$, respectively. Heat transfer only
%	occurs during the transition from state $0\otimes S$ to $1\otimes S$.
%	Work transfer only happens when coupling the input bit to machine's
%	state and when decoupling the output bit from machine's state.
%	}
%\label{CQT}
%\end{figure}

\section{True Random Number Generation}  
\label{sec:TRNG} 

Consider situations in which no random information source is explicitly given
as with RNGs and none is approximated algorithmically as with PRNGs. This
places us in the domain of \emph{true random number generator}s (TRNGs):
randomness is naturally embedded in their substrate physics. For example, a spin one-half
quantum particle oriented in the $z^+$ direction, but measured in $x^+$ and
$x^-$ directions, gives $x^+$ and $x^-$ outcomes with $\sfrac{1}{2}$ and
$\sfrac{1}{2}$ probabilities. More sophisticated random stochastic process
generators employing quantum physics have been introduced recently
\cite{Maho15a,Agha16b,Gu12aa,Tan14aa,Agha16aa,Riech16a}. TRNGs have also been
based on chaotic lasers \cite{Uchi08a,Kant10aa}, metastability in electronic
circuits \cite{Kinn02a,Toku08a}, and electronic noise \cite{Epst03a}. What
thermodynamic resources do these TRNGs require? We address this here via one
general construction.

\begin{figure}
\includegraphics[width=1\linewidth]{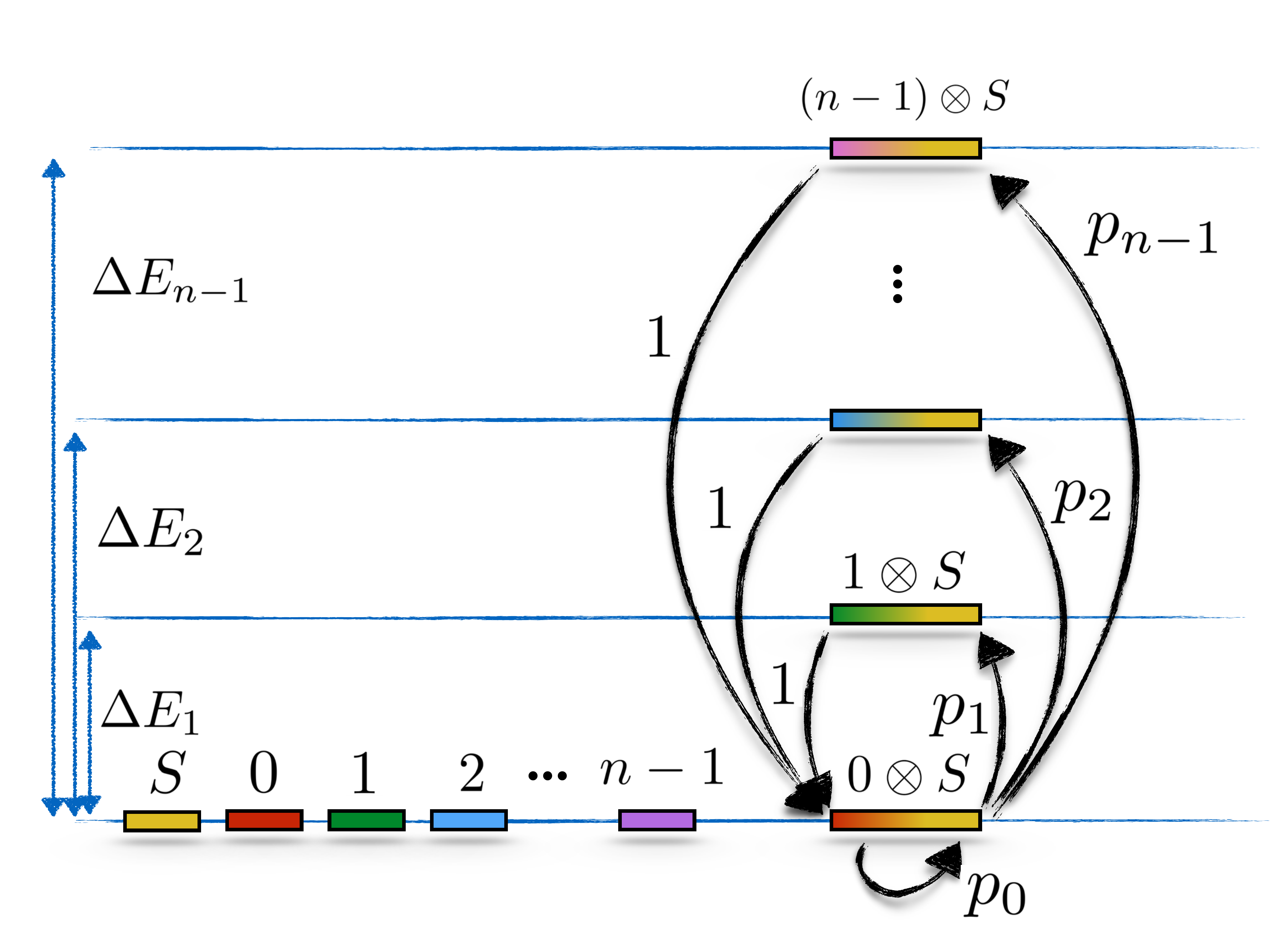}
\caption{True general-distribution generator:  Emit random samples from an
	arbitrary probability distribution $\{p_i\}$, $i = 0, \ldots, n-1$ where $p_1$ to $p_{n-1}$ sorted from large to small. It
	has one internal state $S$ and inputs and outputs can be $0$, $1$,
	..., $n-1$. All states have energy zero. The joint states $i\otimes S$
	for $ i \neq 0$ have nonzero energies $\Delta E_i$. Heat is transferred
	only during the transition from state $0 \otimes S$ to states $i \otimes
	S$. Work is transferred only during coupling the input bit to machine's
	state and decoupling the output bit from machine's state.
	}
\label{GCQT}
\end{figure}

\paragraph*{True General-Distribution Generator:} Consider the general case
where we want to generate a sample from an arbitrary probability distribution
$\{p_i\}$. Each time we need a random sample, we feed in $0$ and the TRNG
returns a random sample. Again, the input is a long sequence $0$s and, as a
consequence, $h=0$. We also have $h^{\prime} = \H[\{p_i\}]$ and $\Lhat = 1$.
Equation~(\ref{DQBound}) puts a bound on the dissipated heat and input work:
$\QLB = -\kB T \ln 2 \H[\{p_i\}]$. Notice here that $\QLB$ is a negative
quantity. This is something that, as we showed above, can never happen for RNG
algorithms since they all are heat-dissipation positive: $\QLB > 0$. Of
course, $\QLB$ is only a lower bound and $\Delta Q$ may still be positive.
However, negative $\QLB$ opens the door to producing work from heat instead
of turning heat to dissipative work---a functioning not possible for RNG
algorithms.

Figure~\ref{GCQT} shows one example of a physical implementation. The machine
has a single state $S$ and the inputs and outputs come from the symbol set $\{
0,1, \cdots, n-1\}$, all with zero energies. The system is designed so that the
joint state $0 \otimes S$ has zero energy and the joint states $i\otimes S$, $i
> 0$, have energy $\Delta E_i$. Recall that every time we need a random sample
we feed a $0$ to the TRNG machine. Feeding $0$ has no energy cost, since the
sum of energies of states $0$ and $S$ is zero and equal to the energy of the
state $0 \otimes S$. Then, putting the system into contact with a thermal
reservoir, we have stochastic transitions between state $0\otimes S$ and the
other states $i\otimes S$. Tuning the $i\otimes S \to 0\otimes S$ transition
probabilities in a fixed time $\tau$ to $1$ and assuming detailed balance, all
the other transition probabilities are specified by the $\Delta E_i$s and,
consequently, for all $i \in \{1, 2, \cdots, n-1\}$, we have
$p_i=\exp{(-\beta\Delta E_i)}$.

The design has the system start in the joint state $0\otimes S$ and after time
$\tau$ with probability $p_i$ it transitions to state $i\otimes S$. Then the
average heat transferred from the system to the thermal reservoir is $-
\sum_{i=1}^{n-1} p_i \Delta E_i$. Now, independent the current state $i\otimes
S$, we decouple the machine state $S$ from the target state $i$. The average
work we must pump into the system for this to occur is:
\begin{align*}
\Delta {W} =  -\sum_{i=1}^{n-1} p_i \Delta E_i
  ~.
\end{align*}
This completes the TRNG specification. In summary, the average heat $\Delta Q$
and the average work $\Delta {W}$ are the same and equal to $\sum_{i=1}^{n-1}
p_i \Delta E_i$.

Replacing $\Delta E_i$ by $-\kB T\ln p_i$ we have:
\begin{align}
\Delta Q = \kB T\sum_{i=1}^{n-1} p_i \ln p_i < 0
  ~,
\label{QTRNG}
\end{align}
which is consistent with the lower bound $-\kB T \ln 2 \H[\{p_i\}]$ given
above. Though, as noted there, a negative lower bound does not mean that we can
actually construct a machine with negative $\Delta Q$, in fact, here is one
example of such a machine. Negative $\Delta Q$ leads to an important physical
consequence. The operation of a TRNG is a heat-consuming and work-producing
process, in contrast to the operation of an RNG. This means not only are the
random numbers we need being generated, but we also have an engine that
absorbs heat from thermal reservoir and converts it to work. Of course, the amount of work depends on the distribution of interest. Thus, TRNGs are
a potential win-win strategy. Imagine that at the end of charging a battery,
one also had a fresh store of random numbers.

Let's pursue this further. For a given target distribution with $n$ elements,
we operate $n$ such TRNG machines, all generating the distribution of interest.
Any of the $n$ elements of the given distribution can be assigned to the
self-transition $p_0$. This gives freedom in our design to choose any of the
elements. After choosing one, all the others are uniquely assigned to $p_1$ to
$p_{n-1}$ from largest to smallest. Now, if our goal is to pump-in less heat
per sample, which of these machines is the most efficient? Looking closely at
Eq. (\ref{QTRNG}), we see that the amount of heat needed by machine $j$ is
proportional to $\H(\{p_i\}) - |p_j \log_2 p_j|$. And so, over all the
machines, that with the maximum  $|p_j \log_2 p_j|$ is the minimum-heat
consumer and that with minimum $|p_j \log_2 p_j|$ is the maximum-work producer.

Naturally, there are alternatives to the thermodynamic transformations used in
Fig.~\ref{GCQT}. One can use a method based on spontaneous irreversible
relaxation. Or, one can use the approach of changing the Hamiltonian
instantaneously and changing it back quasistatically and isothermally
\cite{Parr15aa}.

Let's close with a challenge. Now that a machine with negative $\Delta Q$ can
be identified, we can go further and ask if there is a machine that actually
achieves the lower bound $\QLB$. If the answer is yes, then what is that
machine? We leave the answer for the future.

\section{Conclusion} 
\label{sec:Conclusion} 

Historically, three major approaches have been employed for immediate random
number generation: RNG, PRNG, and TRNG. RNG itself divides into three
interesting problems. First, when we have an IID source, but we have no
knowledge of the source and the goal is to design machinery that generates
an unbiased random number---the von Neumann RNG. Second, when we have a known
IID source generating a uniform distribution and the goal is to invent a
machine that can generate any distribution of interest---the Knuth and Yao RNG.
Third, we have the general case of the second, when the randomness source is
known but arbitrary and the goal is to devise a machine that generates another
arbitrary distribution---the Roche and Hoshi RNG. For all these RNGs the
overarching concern is to use the minimum number of samples from the input
source. These approaches to random number generation may seem rather similar
and to differ only in mathematical strategy and cleverness. However, the
thermodynamic analyses show that they make rather different demands on their
physical substrates, on the thermodynamic resources required.

We showed that all RNG algorithms are heat-consuming, work-consuming processes.
In contrast, we showed that TRNG algorithms are heat-consuming, work-producing
processes. And, PRNGs lie in between, dissipation neutral ($\Delta Q = 0$) in
general and so the physical implementation determines the detailed
thermodynamics. Depending on available resources and what costs we want to pay,
the designer can choose between these three approaches.

The most thermodynamically efficient approach is TRNG since it generates both
the random numbers of interest and converts heat that comes from the thermal
reservoir to work. Implementing a TRNG, however, also needs a physical system
with inherent stochastic dynamics that, on their own, can be inefficient
depending on the resources needed. PRNG is the most unreliable method since it
ultimately produces periodic sequences instead of real random numbers, but
thermodynamically it potentially can be efficient. The RNG approach, though,
can only be used given access to a randomness source. It is particularly useful
if it has access to nearly free randomness source. Thermodynamically, though,
it is inefficient since the work reservoir must do work to run the machine, but
the resulting random numbers are reliable in contrast to those generated vis a
PRNG.

To see how different the RNG and TRNG approaches can be, let's examine a
particular example assuming access to a weakly random IID source with bias $p
\ll 1$ and we want to generate an unbiased sample. We can ignore the randomness
source and instead use the TRNG method with the machine in Fig.~\ref{GCQT}.
Using Eq.~(\ref{QTRNG}) on average to produce one sample, the machine absorbs
$|\kB T ~p \ln p|\approx 0$ heat from heat reservoir and turn it into work.
Since the required work is very small, this approach is resource neutral,
meaning that there is no energy transfer between reservoir and machine. Now,
consider the case when we use the RNG approach---the von Neumann algorithm. To
run the machine and generate one symbol, on average the work reservoir needs
provide work energy to the machine. This thermodynamic cost can be infinitely
large depending on how small $p$ is. This comparison highlights much different
the random number generation approach can be and how is useful depends on the
available resources.

The thermodynamic analysis of the main RNG strategies suggests a number of
challenges. Let's close with several brief questions that hint at several
future directions in the thermodynamics of random number generation. Given that
random number generation is such a critical and vital task in modern computing,
following up on these strike us as quite important. First, is Szilard's Engine
\cite{Szil29a} a TRNG? What are the thermodynamic costs in harvesting
randomness? A recent analysis appears to have provided the answers
\cite{Boyd14b} and anticipates TRNG's win-win property.  Second, the randomness
sources and target distributions considered were rather limited compared to the
wide range of stochastic processes that arise in contemporary experiment and
theory. For example, what about the thermodynamics of generating $1/f$ noise
\cite{Pres78a}? Nominally, this and other complex distributions are associated
with infinite memory processes \cite{Marz15a}. What are the associated
thermodynamic cost bounds? Suggestively, it was recently shown that
infinite-memory devices can actually achieve thermodynamic bounds
\cite{Boyd16d}.  Third, the random number generation strategies considered here
are not secure.  However, cryptographically secure random number generators
have been developed \cite{East15a}. What type of physical systems can be used
for secure TRNG and which are thermodynamically the most efficient? One
suggestion could be superconducting nanowires and Josephson junctions near
superconducting critical current \cite{Folt15aa}. Fourth, what are the
additional thermodynamic costs of adding security to RNGs? Finally, there is a
substantial quantum advantage when compressing classical random processes
\cite{Maho15a}. What are the thermodynamic consequences of using such quantum
representations for RNGs?\\

%\section{Energy Bound with Conservation of Bits}
%
%\begin{align*}
%\frac{Q}{n} \geq \frac{\kB T \ln 2}{n}
%  \big( \H[X_{0:n}] - \H[R_{0:m}] - \H[T_{0:n-m}]\big)
%  ~.
%\end{align*}
%
%This can be written as:
%\begin{align*}
%\frac{Q}{n} \geq \kB T \ln 2 &\big(
%  \frac{\H[X_{0:n}]}{n}
%  -\frac{\H[R_{0:m}]}{m} \left( \frac{m}{n} \right)\\
% &-\frac{\H[T_{0:n-m}]}{n-m} \left( \frac{n-m}{n} \right)
%  \big)
%  ~.
%\end{align*}
%
%\begin{align*}
%\frac{Q}{n} \geq \kB T \ln 2 &\big(
%  \H(\{p_i\})
%  -\frac{\H(\{q_i\})}{L^*} 
% -(\frac{L^*-1}{L^*})\H(\{\gamma_i\}) \big)
%  ~.
%\end{align*}
%
%\begin{align*}
%\Delta Q \geq \kB T \ln 2 &\big(
%  L^*\H(\{p_i\})
%  -\H(\{q_i\})
% -(L^*-1)\H(\{\gamma_i\}) \big)
%  ~.
%\end{align*}
%where 
%\begin{align*}
%\gamma_i = (\frac{L^*}{L^*-1})p_i -  (\frac{1}{L^*-1})q_i
%\end{align*}
%

\vspace{-0.1in}
\section*{Acknowledgments}
\label{sec:acknowledgments}
\vspace{-0.1in}

We thank A. Aghamohammadi, M. Anvari, A. B. Boyd, R. G. James, M. Khorrami, J.
R. Mahoney, and P. M. Riechers for helpful discussions. JPC thanks the Santa Fe
Institute for its hospitality during visits as an External Faculty member. This
material is based upon work supported by, or in part by, the John Templeton
Foundation and U. S. Army Research Laboratory and the U. S. Army Research
Office under contracts W911NF-13-1-0390 and W911NF-13-1-0340.

\vspace{-0.1in}
\bibliography{chaos,torngref}

\end{document}